\documentclass{aa}

\usepackage{graphicx}
\usepackage{txfonts}
\usepackage{natbib}     %A&A reference needed usepackage
\usepackage{xcolor} %for color text
\bibpunct{(}{)}{;}{a}{}{,} % for citation style,to follow the A&A style

\begin{document}

   \title{The effect of the stellar absorption line centre-to-limb variation on exoplanet transmission spectrum observations}

   \author{F. Yan
          \inst{1}
          \and
          E. Pall\'e
          \inst{2,3}
          \and
          R. A. E. Fosbury
          \inst{4}
          \and
          M. G. Petr-Gotzens
          \inst{4}
          \and
          Th. Henning
          \inst{1}
                }

   \institute{Max Planck Institute for Astronomy, K\"onigstuhl 17, 69117, Heidelberg, Germany\\
\email{fyan@mpia.de}
                \and
                        Instituto de Astrof\'isica de Canarias, C/ v\'ia L\'actea, s/n, 38205 La Laguna, Tenerife, Spain
        \and                    
                        Dpto. de Astrof\'isica, Universidad de La Laguna, 38206 La Laguna, Tenerife, Spain
                \and
                        European Southern Observatory, Karl-Schwarzschild-Str. 2, 85748 Garching bei M\"unchen, Germany
\\      
         }
        \date{Received -----; accepted -----}
  % \date{Received September 15, 1996; accepted March 16, 1997}

% \abstract{}{}{}{}{}
% 5 {} token are mandatory

  \abstract
 {Transit spectroscopy is one of the most commonly used techniques for exoplanet atmosphere characterisation. This technique has been used to detect ionized and neutral species in exoplanet atmospheres by comparing the observed stellar lines in and out of transit. The centre-to-limb variation (CLV) of the stellar lines across the stellar disk is an important effect for transmission spectroscopy, since it results in a change of stellar line depth when the planet transits different parts of the stellar disk. 
 We reanalyse the transit data of HD 189733b taken with the HARPS spectrograph to study the CLV effect during transit. The transmission light curve of the \ion{Na}{i} D line so obtained shows a clear imprint of the CLV effect. We use a one-dimensional non-LTE stellar spectral model to simulate the CLV effect. After applying the correction, the measurement of the \ion{Na}{i} absorption in the atmosphere of HD 189733b becomes better determined. 
We compare the CLV effect of HD 189733b to that of HD 209458b. The CLV effects are different for these two benchmark planetary systems and this is attributed to their different stellar effective temperatures and transit impact parameters.
We then explore the general CLV effect that occurs during exoplanet transits.
Normally, a star with a lower effective temperature exhibits a stronger CLV effect and its CLV feature extends over a relatively broad wavelength range. The transit impact parameter (b) describes the transit trajectory on the stellar disk and thus determines the actual manifestation of the CLV effect. We introduce a b-diagram which describes the behavior of the CLV effect as the function of different impact parameters.
With improving observational precision, a careful modeling and correction of the CLV effect is necessary for exoplanet atmosphere characterisation using transit spectroscopy.
 }

   \maketitle

%
%________________________________________________________________

\section{Introduction}
Atmospheric characterisation is a rapidly expanding branch of exoplanet research.  One of the most commonly used methods for this is transit spectroscopy.
Shortly after the first discovery of a transiting exoplanet -- HD 209458b by \cite{Charbonneau2000}, the first detection of an exoplanet atmosphere was achieved by \cite{Charbonneau2002} with the Hubble Space Telescope (HST). This revealed atomic sodium absorption in its atmosphere. \cite{Sing2008} reanalysed the HST data and obtained an optical transmission spectrum which shows a broad \ion{Na}{i} absorption profile.  \citet[hereafter S2008]{Snellen2008} confirmed the \ion{Na}{i} absorption in \object{HD 209458b} using transit spectra taken with the High Dispersion Spectrograph on the Subaru telescope -- this is one of the first detections of exoplanet atmospheres from the ground.  
In addition to the atomic sodium, other species are also detected in the HD 209458b atmosphere, such as atomic hydrogen, oxygen and carbon \citep{Vidal-Madjar2003, Vidal-Madjar2004}.

Apart from HD 209458b, another extensively studied exoplanet is HD 189733b. 
\cite{Redfield2008} presented the first atmospheric detection for this planet. The detection, made with the High Resolution Spectrograph on the Hobby-Eberly Telescope, reveals a \ion{Na}{i} absorption feature. \cite{Pont2013} showed the full transmission spectrum from the UV to infrared wavelengths and revealed sodium and potassium absorption as well as a strong Rayleigh scattering. Recently, \citet[hereafter W2015]{Wyttenbach2015} obtained the spectrally resolved \ion{Na}{i} transmission spectrum of HD 189733b using the High Accuracy Radial velocity Planet Searcher (HARPS) mounted on the ESO-3.6m telescope \citep{Mayor2003}. These HARPS data are also used by \cite{Louden2015} for the study of wind and rotation, and by \cite{Barnes2016} for studying excess transit absorption at  H$\alpha$ and  \ion{Ca}{ii} H\&K line cores.

Transit spectroscopy not only plays an important role in the discovery of atmospheric species, but also has great potential for characterising the physical condition of atmospheres. 
For example, \cite{Snellen2010} detected a strong wind from the day- to night-side with a speed of 2 $\mathrm{km s^{-1}}$ for HD 209458b using the cross-correlation method of the CO lines. \cite{Brogi2016} applied a similar method to HD 189733b and detected a wind speed of 1.7 $\mathrm{km s^{-1}}$ using the CO and $\mathrm{H_2O}$ lines. \cite{Louden2015} also detected the wind speed of HD 189733b but with the Na D lines.

The transit spectroscopy technique utilises the stellar flux as the background source which penetrates the planetary atmosphere during transit, thus the non-uniformity of the stellar flux across the stellar disk is important for this technique. For instance, the stellar limb-darkening is a crucial parameter. Normally, the broadband differential limb-darkening is well treated in exoplanet studies, however, the stellar lines have a different limb-darkening from the continuum. This behaviour is called the center-to-limb variation (CLV) effect, i.e. the normalised stellar line shows a variation in the line profile from the center to limb across the stellar disk. 
The CLV effect can affect the transit spectroscopy studies which use narrow-band light curve measurements as well as spectrally-resolved measurements including the cross-corelation method.

For the solar Fraunhofer lines,  there are prominent variations in both line intensity and profile across the solar disk \citep{Athay1972}. 
In our previous work \citep{Yan2015}, we observed the Earth transiting the Sun using a lunar eclipse to obtain the transmission spectrum of the Earth's atmosphere. We found that the CLV effect in solar lines results in large residuals in the transmission spectrum obtained and that the CLV effect can lead to the false detection of atmospheric species. For example, the detection of  \ion{Na}{i} was claimed in some other lunar eclipse observations \citep{Palle2009, Vidal-Madjar2010, Arnold2014}, but was demonstrated to be the result of the CLV effect rather than of actual planetary absorption.

The importance of the CLV effect has been realized in different studies when investigating transmission spectra. 
For example, \cite{Charbonneau2002} and \cite{Redfield2008} both considered the CLV effect when measuring the sodium in exoplanet atmospheres and concluded that the CLV effect contribution is much smaller than the observed \ion{Na}{i} absorption for their observations.
%However, for stellar lines that have significant CLV effects or if the observation is performed at high spectral resolution, this effect could become important.
However, with the continuous improvements in instrumentation, observing strategies and data reduction methods, the CLV effect can no longer be neglected.
Its manifestation differs for different planetary systems and for different stellar lines. In some cases, the CLV effect can be very strong, becoming crucial for transmission studies.

\citet[hereafter C2015]{Czesla2015} studied the CLV effect and used UVES/VLT data of HD 189733b's transit as an example.
However, their data are affected by a flare event and the line core is not presented in their paper. The line core  is very important for exoplanet characterisation because the planetary absorption is most prominent there. Here, we will present the CLV effect of HD 189733b using another data set from HARPS observations and focus on the CLV at the line core.

Our paper is organised as follows.
In Sect.2, we describe the stellar spectral model for the CLV effect. In Sect.3, we present the HARPS data of the HD 189733b transit and study how the \ion{Na}{i} transmission spectrum is affected by the CLV effect.
In Sect.4, we compare the CLV effects of the two extensively studied exoplanet systems, HD 189733b and HD 209458b, in order to illustrate its manifestation in different exoplanet systems.
In Sect.5, we present the general CLV effect during exoplanet transit and how the CLV effect varies with different planetary and stellar parameters, especially the transit impact parameter and the stellar effective temperature.

%_____________________________________________________________
\section{The CLV model for exoplanet transit}
The model comprises two parts. Firstly we simulate the synthetic stellar spectrum across the stellar disk, and then integrate the un-obscured stellar spectrum to obtain the stellar spectrum visible during transit.

\subsection{Stellar spectrum as a function of limb angle}
The stellar spectrum is modeled using the Spectroscopy Made Easy (SME) tool \citep{Piskunov2016}. We use the MARCS \citep{Gustafsson2008} stellar atmosphere model which is included in the distribution of the SME package. The non-local thermodynamic equilibrium (non-LTE) effect can also be included in the SME calculation, and this is done by applying the appropriate departure coefficients. 
For example, the \ion{Na}{i}  departure coefficients are from \cite{Lind2011} and \cite{Mashonkina2008}. 
As presented in \cite{Piskunov2016}, the SME non-LTE calculation using the MARCS atmosphere grid reproduces well the line profiles of solar \ion{Na}{i} D lines.
We use the line lists from VALD database \citep{Ryabchikova2015}.

The stellar spectra are calculated at 21 limb angles ($\theta$, the angle between the normal to the stellar surface and the line of sight). The corresponding $\mu=\mathrm{cos(}\theta)$ values range from 0 to 1 with a step of 0.05. When calculating the spectrum at the stellar edge, we use $\mu=0.001$ instead of  $\mu=0$ to avoid numerical problems (C2015).
According to the simulation, the strong stellar lines are always deeper at the stellar disk center than at the stellar limb.

\subsection{Synthetic spectrum during transit}
In order to obtain the synthetic spectrum of the integrated stellar disk, we divide the disk into elements with a size of $0.01\mathrm{R_s} \times0.01\mathrm{R_s}$. Each of these elements has a $\mu$ value, and so its spectrum is linearly interpolated from the previously calculated spectra at 21 $\mu$ values.
The synthetic spectrum during transit is calculated by integrating all the surface elements which are not obscured by the planet.

We synthesize two different kinds of spectra: the complete spectrum and the continuum. In this way, we can obtain the normalised spectrum by dividing the complete spectrum by the continuum. The CLV calculation is then performed based on these normalied spectra. 

\subsection{Other effects}
In addition to the CLV, there are other effects that also change the observed line depth during transit.

\textit{\textbf{(1) }Rossiter-McLaughlin effect} \\
The CLV model in our work does not include the Rossiter-McLaughlin (RM) effect, but this will only have a very small effect on our calculations. The RM effect means that the observed stellar line profile and Radial Velocity (RV) change during transit because of the stellar rotation \citep{Rossite1924, McLaughlin1924}.
For a star like HD 189733 with a rotational velocity of $V\mathrm{sin}I=3.1$ $\mathrm{km s^{-1}}$ \citep{Triaud2009}, the stellar line shape change due to the RM effect is relatively weak.
Besides, when using transit spectroscopy to detect the transmission spectrum, normally the RV differences between the observed stellar spectra are corrected by shifting the spectra to a common velocity before analysis. For the transmission light curve method we employed in the following analysis in Section 3 , the spectra are integrated over a narrow band and thus the line distortion due to the RM effect has a relatively weak effect on our result.
Therefore, we do not include the RM effect in the model.

However, we emphasize here that the RM effect is important for spectrally resolved transmission spectroscopy. 
For instance, \cite{Louden2015} showed that the blue-shifted Na absorption of HD 189733b measured by \cite{Wyttenbach2015} is likely due to the uncorrected RM effect. \cite{Brogi2016} also showed that the uncorrected RM effect will produce a CO signal at 2.3 $\mathrm{\mu m}$ dominated by the residual of stellar absorption, rather than by the true planet signal.
In addition to the RM effect, the CLV effect is also very important for these spectrally resolved studies as it changes the observed stellar line profile.

\textit{\textbf{(2) }Planet orbital velocity} \\
The planet orbital RV varies during transit, for example, the RV shift of HD 189733b during transit is $\pm 16 ~ \mathrm{km/s}$. The change of orbital RV causes the shift of the planetary absorption line across the stellar line profile, thus the actual absorbed flux by the planetary atmosphere varies with the orbital RV.  For example, when the orbital RV equals zero, the planetary absorption line is mainly centered at the stellar line core where the flux is low, thus the flux absorbed by the planet is relatively small; when the absorption line shifts away from the stellar line core, the flux absorbed by the planet becomes relatively larger. This orbital RV effect is mentioned in \cite{Albrecht2008PhD} and \cite{Snellen2011}. 
More recently, \cite{Khalafinejad2016} observed the light curve of a HD 189733b transit due to this orbital RV effect using the UVES/VLT data. 
Generally, the flux change due to the orbital RV is relatively small and thus it is not included in our CLV model. 
However, depending on the stellar and planetary absorption line profiles and the chosen passband, the planet orbital RV can introduce additional noise into the observed transmission light curve.

%%%%

%_____________________________________________________________
\section{The CLV effect in the HD 189733b data}
 HARPS observed HD189733b transits during several nights. These data have been extensively studied by other authors, e.g. \cite{Wyttenbach2015}, \cite{Louden2015}, \cite{Di-Gloria2015} and  \cite{Barnes2016}.
 Three datasets taken at different nights are used by W2015 and \cite{Barnes2016} for studying the \ion{Na}{i} D, H$\alpha$ and  \ion{Ca}{ii} H\&K lines. %Table 1 shows the details of the datasets.
 In this work, we use two datasets: from Night 1 (2006-09-07) and Night 3 (2007-08-28). The Night 2 (2007-07-19) data are not used because they lack sufficient baseline observations before transit.

The standard data reduction is performed using the HARPS Data Reduction Software. The reduced spectrum is one-dimensional, instrument response (blaze) and Baycentric Earth Radial Velocity corrected. The wavelength of the reduced spectrum is in the solar system barycenter rest frame.

For telluric correction, we use a similar method to  W2015. The telluric correction mainly corrects the $\mathrm{H_2O}$ lines in the \ion{Na}{i} D wavelength region. Firstly, we build a telluric reference spectrum for each night using the linear regression method with the out-of-transit spectra (c.f. W2015 and \cite{Astudillo-Defru2013} for more details). Secondly, we fit the strong  $\mathrm{H_2O}$ lines in the obtained telluric reference spectra with the corresponding lines in each observed spectrum, so that a telluric transmission spectrum is obtained from the fit solution. Each observed spectrum is then divided by its corresponding telluric transmission spectrum to remove the telluric lines.

%
%                                             Simple A&A Table
%-----------------------------------------------------------
\begin{table}
\caption{Parameters of the HD 189733b system. The transit epoch and the period are from \cite{Agol2010} and all the other values are from \cite{Torres2008}. }             % title of Table
\label{parameter-HD189}      % is used to refer this table in the text
\centering                          % used for centering table
\begin{tabular}{l c c  }        % centered columns (4 columns)
\hline\hline                 % inserts double horizontal lines
%~	&	HD189733b &	HD209458b & ~ & ~ \\     % table heading
	Parameter & Symbol [unit] & Value   \\
	\hline                       % inserts single horizontal line
	\textit{The star} & ~ & ~ \\
 		Effective temperature		& $T_\mathrm{eff} [K]$	&	5040 $\pm$ 50   \\
        Radius 	& $R_\mathrm{s}$ [$R_\odot$]	&	0.756 $\pm$ 0.018   \\
        Mass	& $M_\mathrm{s}$	[$M_\odot$]&	0.806  $\pm$ 0.048  \\
        Surface gravity		& $\mathrm{log}( g [\mathrm{cm s^{-2}}])$	&	4.587$^{+0.014}_{-0.015}$  \\
        Metallicity	&  [M/H] [dex]	&	-0.03 $\pm$ 0.08  \\
        ~ & ~ & ~  \\
        \textit{The planet} & ~ & ~  \\
        Radius ratio 	& $R_\mathrm{p}/R_\mathrm{s}$	& 0.15463 $\pm$ 0.00022   \\
       % Mass	& $M_\mathrm{p}$	[$M_\mathrm{J}$] 	&	-0.08   \\
        Orbital semi-major axis & $a$	[$R_\mathrm{s}$]	 &	8.81 $\pm$ 0.06  \\
        Orbital inclination & $i$ [degrees] 	&	85.58 $\pm$ 0.06    \\
        Orbital period & $P$ [days]	&	2.21857567  \\
        Transit impact parameter & $b$ [ $R_\mathrm{s}$]	&	0.680 $\pm$ 0.005   \\
        Transit epoch (BJD) & $T_\mathrm {0}$ [days]	&	2454279.436714   \\
        
\hline                                   %inserts single line
\end{tabular}
\end{table}
%-----------------------------------------------------------

%
%                                             Simple A&A Table
%-----------------------------------------------------------
\begin{table}
\caption{Parameters of the HD 209458b system. The transit epoch and the period are from \cite{Knutson2007} and all the other values are from \cite{Torres2008}. }           % title of Table
\label{parameter-HD209}      % is used to refer this table in the text
\centering                          % used for centering table
\begin{tabular}{l c c  }        % centered columns (4 columns)
\hline\hline                 % inserts double horizontal lines
%~	&	HD189733b &	HD209458b & ~ & ~ \\     % table heading
	Parameter & Symbol [unit] & Value   \\
	\hline                       % inserts single horizontal line
	\textit{The star} & ~ & ~ \\
 		Effective temperature		& $T_\mathrm{eff} [K]$	&	6065 $\pm$ 50   \\
        Radius 	& $R_\mathrm{s}$ [$R_\odot$]	&	1.155$^{+0.014}_{-0.016}$   \\
        Mass	& $M_\mathrm{s}$	[$M_\odot$]&	1.119  $\pm$ 0.033  \\
        Surface gravity		& $\mathrm{log}( g [\mathrm{cm s^{-2}}])$	&	4.361$^{+0.007}_{-0.008}$  \\
        Metallicity	&  [M/H] [dex]	&	0.00  $\pm$ 0.05 \\
        ~ & ~ & ~  \\
        \textit{The planet} & ~ & ~  \\
        Radius ratio 	& $R_\mathrm{p}/R_\mathrm{s}$	& 0.12086 $\pm$ 0.00010   \\
       % Mass	& $M_\mathrm{p}$	[$M_\mathrm{J}$] 	&	-0.08   \\
        Orbital semi-major axis & $a$	[$R_\mathrm{s}$]	 &	8.76 $\pm$ 0.04  \\
        Orbital inclination & $i$ [degrees] 	&	86.71 $\pm$ 0.05    \\
        Orbital period & $P$ [days]	&	3.52474859   \\
        Transit impact parameter & $b$ [ $R_\mathrm{s}$]	&	0.507 $\pm$ 0.005  \\
        Transit epoch (BJD) & $T_\mathrm {0}$ [days]	&	2452826.628514   \\
        
\hline                                   %inserts single line
\end{tabular}
\end{table}
%-----------------------------------------------------------

%-----------------------------------------------------------
\subsection{Observed transmission light curve of \ion{Na}{i} and the CLV correction}
\begin{figure}
   \centering
   \includegraphics[width=0.50\textwidth]{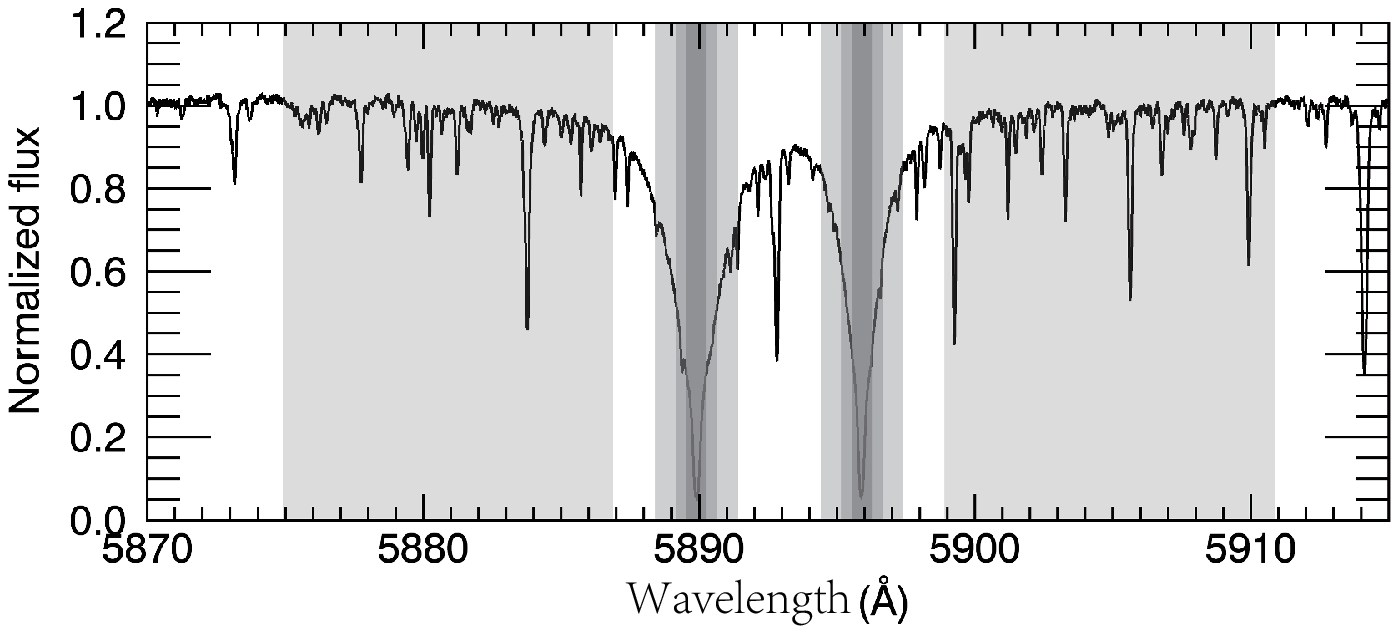}
    \includegraphics[width=0.50\textwidth]{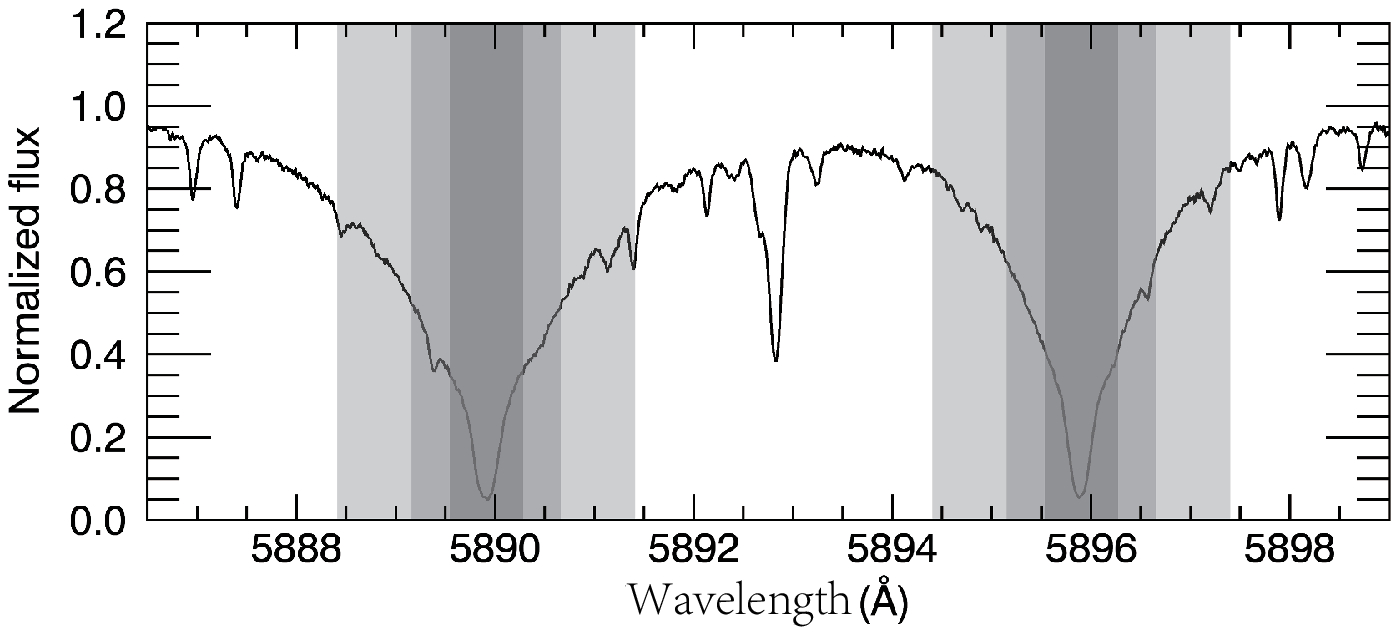}
\caption{
The observed spectrum of HD 189733. 
\textit{The top panel} shows the three bandwidths (0.75$\,\AA$, 1.5$\,\AA$ and 3.0$\,\AA$) used to measure the relative flux of the \ion{Na}{i} D lines. These passbands are centered at the line cores of $\mathrm{D_1}$ and $\mathrm{D_2}$. The reference passbands at the blue and red parts of the \ion{Na}{i} D lines are also shown.
\textit{The bottom panel} is the zoom at the the \ion{Na}{i} D lines. 
}
         \label{observed-spectrum}
   \end{figure}

We choose three bandwidths to calculate the line depth: 0.75$\,\AA$, 1.5$\,\AA$ and 3.0$\,\AA$.  These bands are centered at the line cores of each $\mathrm{D_1}$ and $\mathrm{D_2}$ line. 
The reference passband used is the same as in W2015, i.e. 5874.89 $\sim$ 5886.89$\,\AA$ for the blue part and 5898.89 $\sim$ 5910.89$\,\AA$ for the red part (Fig. \ref{observed-spectrum}). 
The relative flux of  the \ion{Na}{i} D lines is calculated using the ratio between the flux inside the band centered at the line core ($F_\mathrm{cen}$) and the flux inside the reference bands ($F_\mathrm{red}$ and $F_\mathrm{blue}$):
\begin{equation}
      F_\mathrm{line} = \frac{2~F_\mathrm{cen} }{F_\mathrm{red}+F_\mathrm{blue} }
   \end{equation}
We normalise the relative flux to unity for all the out-of-transit $F_\mathrm{line}$ values.
In this way, we can obtain the transmission light curves for $\mathrm{D_1}$ and $\mathrm{D_2}$ separately. The light curves of  $\mathrm{D_1}$ and $\mathrm{D_2}$ are then averaged.

%                                                one column figure
%----------------------------------------------------------- S_vib

\begin{figure*}
   \centering
   \includegraphics[width=0.98\textwidth]{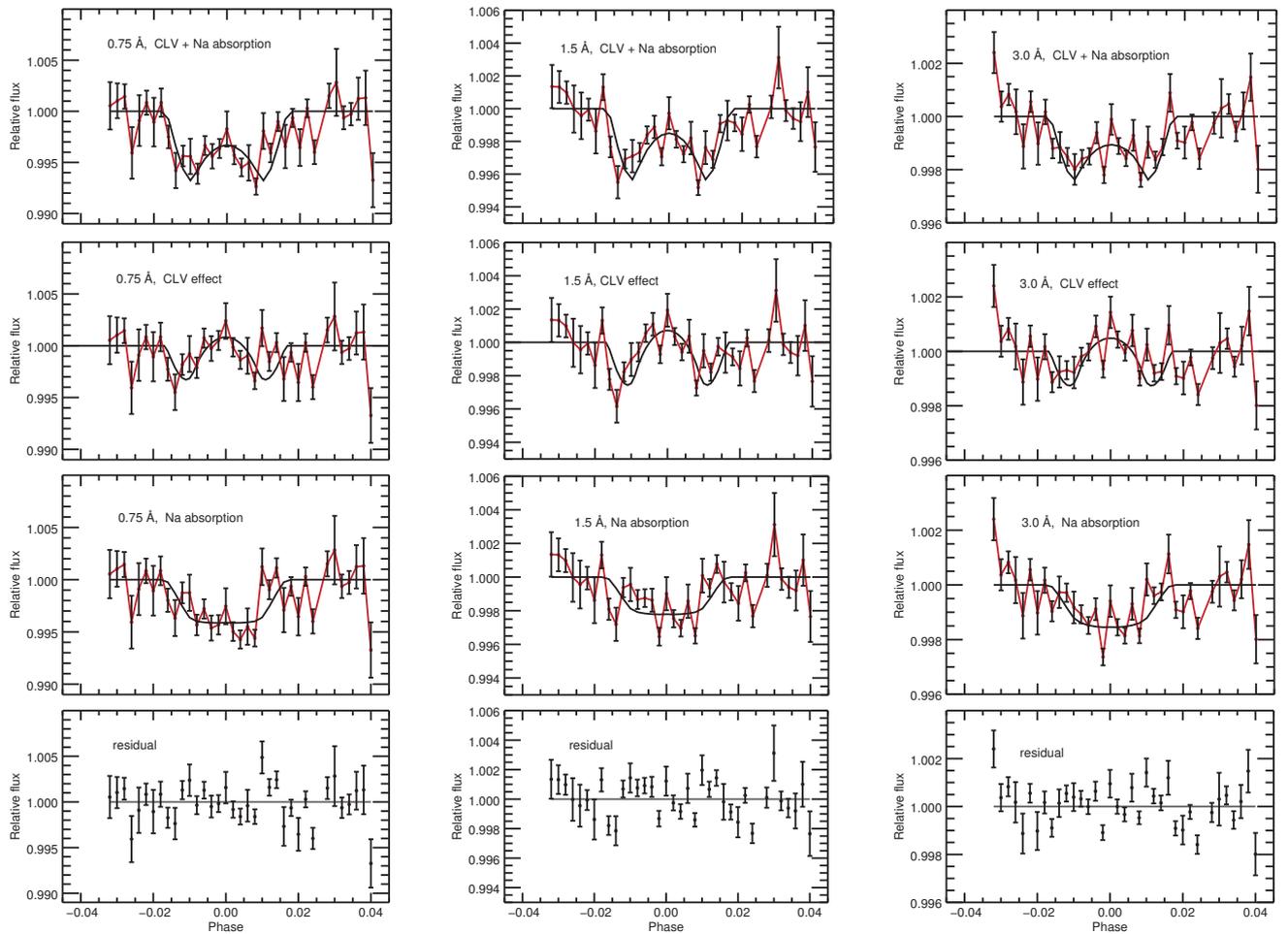}
\caption{
\textbf{First row:} the observed relative flux of the \ion{Na}{i} D lines. This is the final observed \textit{transmission light curve} from our data reduction.
The data are from the HARPS observations of HD 189733b. The data points are the average of $\mathrm{D_1}$ and $\mathrm{D_2}$ lines and are binned every 0.02 phase step. The black line is the modeled transmission light curve which is the combination of the CLV effect and the best-fit \ion{Na}{i} absorption.
\textbf{Second row:} the \textit{CLV light curve}. The black line is the modeled CLV effect and the red line is the observed light curve which is obtained by removing the \ion{Na}{i} absorption model from the observed transmission light curve.
\textbf{Third row:} the \textit{\ion{Na}{i} absorption light curve}. The red line is the observed light curve with the modeled CLV effect removed and the black line is the best-fit  \ion{Na}{i} absorption light curve. 
\textbf{Fourth row:}  the residuals between the observed data (the top row) and the model.
The three columns correspond to the observed flux measured using the band widths of 0.75$\,\AA$, 1.5$\,\AA$ and 3.0$\,\AA$, respectively. The errorbars represent the photon noise within the passband.
} 
         \label{Na-overall}
   \end{figure*}

For each night, we obtain one transmission light curve of the \ion{Na}{i} D lines. The light curves from Night 1 and Night 3 are then combined and re-ordered according to the phase values. We bin the data points with a 0.002 phase step. In each bin, the data points are weighted by their SNRs. The final observed transmission light curves using different passbands are shown in the first row of Fig. \ref{Na-overall} (the red line).

The light curve is the combination of the planetary absorption and the CLV effect. To better characterise the planetary  \ion{Na}{i} absorption, the CLV effect needs to be corrected. 
We use the parameters of the HD 189733 system listed in Table 1 as the input to calculate the relative flux change of the \ion{Na}{i} lines due to the CLV effect. The modeled light curve of the CLV effect is presented as the black line in the second row of Fig. \ref{Na-overall}.

We correct the CLV effect by dividing the observed relative flux light curve with the modeled CLV light curve in order to obtain the actual  \ion{Na}{i} absorption light curve (red line in the third row of Fig. \ref{Na-overall}). The  \ion{Na}{i} absorption depth is then calculated by fitting with a modeled \ion{Na}{i} light curve. The best-fit model of \ion{Na}{i} is plotted as the black line in the third row of Fig. \ref{Na-overall}.

In Fig. \ref{Na-overall}, we also plot the combined model (the combination of the CLV model and the best-fit \ion{Na}{i} absorption) as the black line in the first row. Additionally, by dividing the observed transmission light curve with the best-fit \ion{Na}{i} absorption model, we can obtain the observed CLV effect (the red line in the second row of Fig. \ref{Na-overall}. We emphasise here that the modeled CLV effect light curve is directly from the model and not from the fit procedure. The fit procedure is only performed for the \ion{Na}{i} absorption light curve. Once the best-fit \ion{Na}{i} absorption model is determined, the combined model and the observed CLV light curve are then obtained directly.

%                                                two column figure
%----------------------------------------------------------- S_vib

Table \ref{Table-Na-HD189} presents numeric values of the results. The obtained \ion{Na}{i} absorption depths are 0.414 $\pm$ 0.032\%, 0.222 $\pm$ 0.018\% and 0.155 $\pm$ 0.011\% at the 0.75 $\,\AA$, 1.5 $\,\AA$ and 3.0 $\,\AA$ bands.
Here, the CLV effect is presented as three values: \textbf{mid-transit} (phase=0), \textbf{limb-transit} (where the line depth change is the largest and the planet transits mainly the limb part of the stellar disk, phase$ \approx $0.01 for HD 189733b),  \textbf{overall-transit} (average of all the line depth changes in transit). We introduce this overall-transit value because combining all the in transit spectra is a commonly used method for exoplanet atmosphere detection. Here a positive value represents an absorption feature (i.e. the line depth is deeper compared to out of transit), while a negative value represents an excess in line depth compared to out of transit. The last column of the table is the overall CLV contribution (the ratio of the overall-transit values to the actual  \ion{Na}{i} absorption depth).

\subsection{Conclusions for HD 189733b}
The observed transmission light curve can be well fitted with the combined model (CLV plus \ion{Na}{i} absorption). 
If we fit the light curve with only the \ion{Na}{i} absorption model, then the best-fit absorption depths are 0.512\%, 0.292\% and 0.175\% at the 0.75 $\,\AA$, 1.5 $\,\AA$ and 3.0 $\,\AA$ bands, respectively. These absorption depths are larger than the absorption depths obtained with the combined model, for example, the depth at the 1.5 $\,\AA$ band is larger by a factor of about one third.
The observed light curves fit better with the combined model than with only the \ion{Na}{i} absorption model. 
For example, the reduced chi-square from the best-fit at the 1.5 $\,\AA$ band is 3.3 for the model without CLV effect while the reduced chi-square is 2.6 for the combined model. 
Therefore, we conclude that the model with the CLV effect included fits better with the observed light curve and makes the measurement of \ion{Na}{i} absorption more precise. For the case of HD 189733b, the \ion{Na}{i} absorption depth is overestimated if the CLV effect is not considered, which will result in a considerable overestimation of the \ion{Na}{i} abundance.

Although the CLV value drops with a broader bandwidth, the planetary absorption values also reduce. Thus, the CLV effect compared with the actual planetary absorption does not necessary become less important with an increasing bandwidth. For HD 189733b, the ratios of the overall-transit CLV value to the Na absorption value are similar for the three band widths (c.f. the last column in Table \ref{Table-Na-HD189}). 
Here one should keep in mind that the CLV effect is wavelength dependent. If this correction is not properly taken into account, the obtained wavelength dependent absorption line profile will deviate from its actual shape, influencing the determination of the atmosphere physical conditions (e.g temperature-pressure profile).

%
%                                             Simple A&A Table
%-----------------------------------------------------------
\begin{table*}
\caption{The measured \ion{Na}{i} absorption depth of HD 189733b (CLV effect corrected) and the modeled line depth change due to the CLV effect. The values are obtained using the average of $\mathrm{D_1}$ and $\mathrm{D_2}$ lines. Here, the CLV effect are presented in three values: mid-transit (phase = 0), limb-transit (where the line depth change is the largest) and overall-transit (average of all the line depth changes in transit). }            % title of Table
\label{Table-Na-HD189}      % is used to refer this table in the text
\centering                          % used for centering table
\begin{tabular}{c c c c c c }        % centered columns (4 columns)
\hline\hline                 % inserts double horizontal lines
Passband ($\,\AA$)	&	 Planetary absorption depth (\%)	 &	~ & CLV effect model  (\%) & ~ & CLV overall-transit /  \\     % table heading
	~ & (from this work) & mid-transit & limb-transit & overall-transit & planetary absorption \\
	\hline                       % inserts single horizontal line
 		0.75	& 0.414 $\pm$ 0.032	&	-0.082  & 0.328 & 0.120 & 29\%\\
        1.5 	& 0.222	$\pm$ 0.018  &	-0.071  & 0.257 & 0.091 & 41\%\\
        3.0		& 0.155	$\pm$ 0.011  &	-0.048  & 0.126 & 0.038 & 24\%\\
\hline                                   %inserts single line
\end{tabular}
\end{table*}
%-----------------------------------------------------------

\begin{figure}
   \centering
   \includegraphics[width=0.50\textwidth]{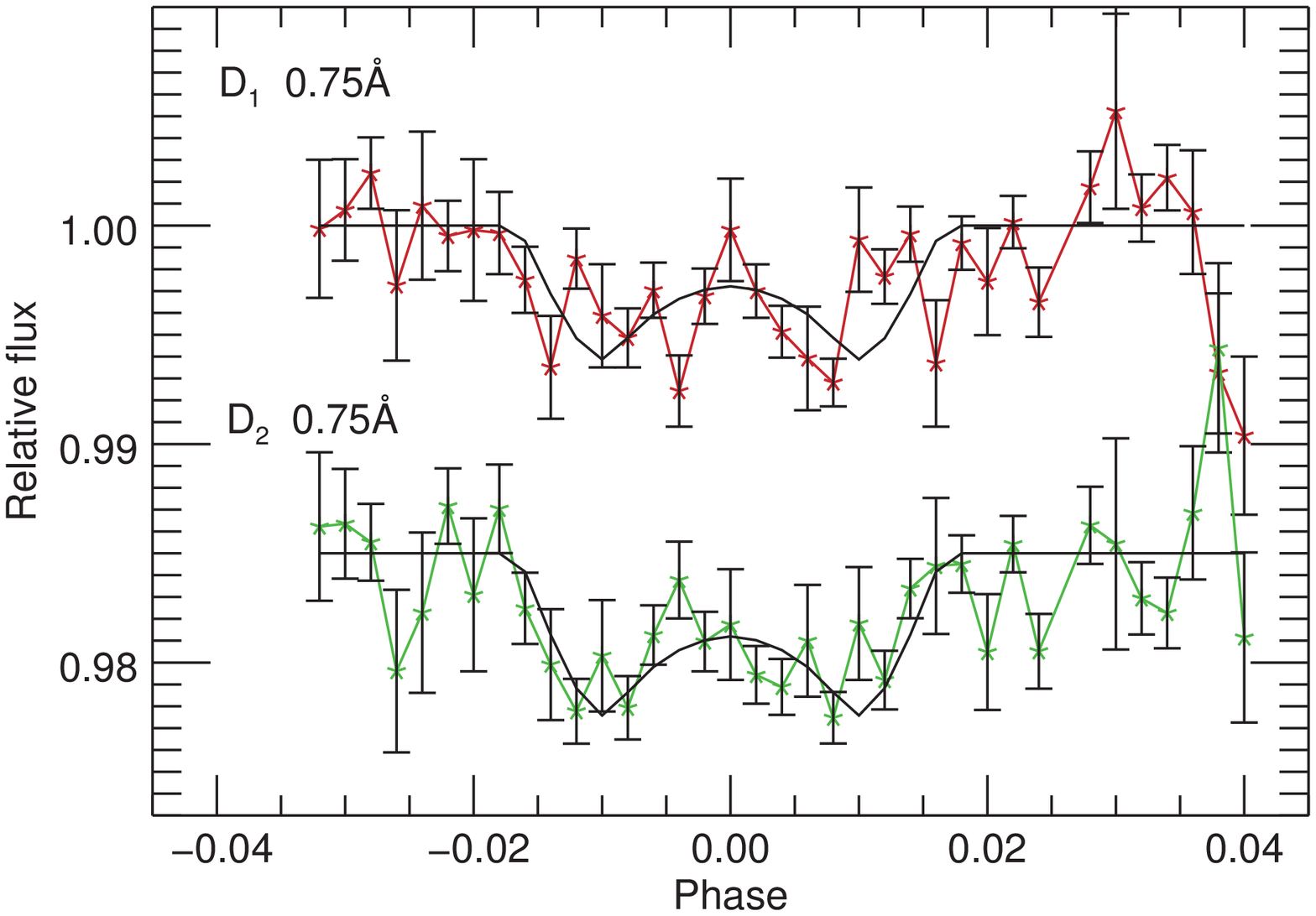}
\caption{The observed transmission light curves for  $\mathrm{D_1}$ and  $\mathrm{D_2}$ measured using a 0.75$\,\AA$ band. The  $\mathrm{D_2}$ curve is shifted down by 0.15.
The data is from HARPS observations of HD 189733b's transit and are binned every 0.02 phase step. The black solid lines are the best-fit combined models (the CLV effect plus the \ion{Na}{i} absorption). The average of these two light curves is the upper-left plot in Fig.~\ref{Na-overall}. }
         \label{D1-D2-compare}
   \end{figure}

The above calculations are for the combination of the \ion{Na}{i} D doublet. We also provide separate results for $\mathrm{D_1}$ and $\mathrm{D_2}$ in Fig.~\ref{D1-D2-compare}. In general, the $\mathrm{D_1}$ and $\mathrm{D_2}$ lines have similar \ion{Na}{i} absorption depth and CLV feature. Their difference is that both the planetary absorption and the CLV effect for  $\mathrm{D_2}$ are slightly stronger than for $\mathrm{D_1}$. 
For example, for the  0.75$\,\AA$ band, the CLV effect (overall-transit value) is 0.125\% for $\mathrm{D_2}$ while it is 0.115\% for $\mathrm{D_1}$, and the planetary absorption depth is 0.470\% for $\mathrm{D_2}$ while it is 0.359\% for $\mathrm{D_1}$.

%                                                one column figure
%----------------------------------------------------------- S_vib

\subsubsection{CLV for \ion{Mg}{i}  $\mathrm{b_1}$ line}
\begin{figure*}
   \centering
   \includegraphics[width=0.98\textwidth]{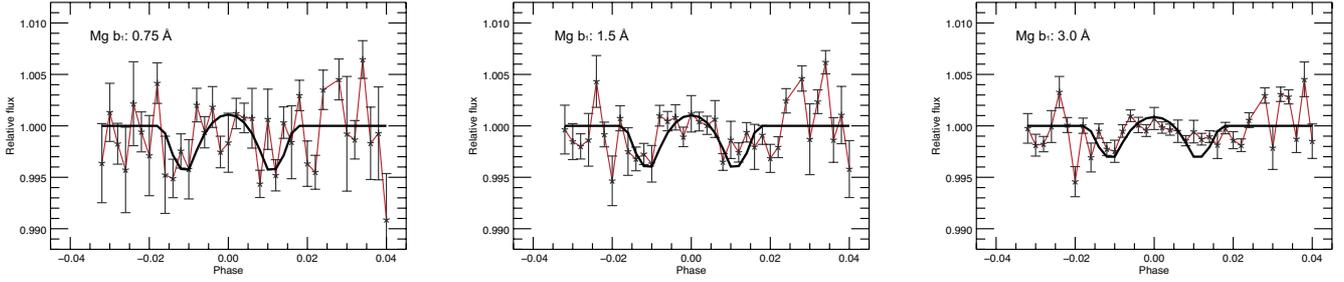}
\caption{The observed light curves for the \ion{Mg}{i}  $\mathrm{b_1}$ line (red lines). The black line is the CLV model. }
         \label{Mg-b1.eps}
   \end{figure*}

Unlike the \ion{Na}{i} D resonance lines, there is normally no planetary absorption at \ion{Mg}{i} b lines. This implies that we can use the \ion{Mg}{i} b line as the compararison line to confirm the CLV effect. Fig.~\ref{Mg-b1.eps} is the observed result for the \ion{Mg}{i}  $\mathrm{b_1}$ line along with the CLV model. The line center is at 5183.59 $\AA$ (in the solar system barycentric rest frame) and the reference bands are 5175.00 $\sim$ 5180.00 $\AA$ and 5187.00 $\sim$ 5191.00 $\AA$. We do not apply the telluric correction for \ion{Mg}{i}  $\mathrm{b_1}$ because the telluric absorption is relatively weak in this wavelength range.
This figure shows that the observed light curve of the \ion{Mg}{i} b line is consistent with the CLV model.

\subsubsection{Comparison with previous results from HARPS observations }
When comparing the results with those in W2015, we find that our \ion{Na}{i} absorption depths (without the CLV correction) are larger than theirs. For example, at the 0.75 $\,\AA$ bandwidth, the average depth of Night 1 and Night 3 in W2015 is 0.40\% while it is 0.512\% in our work. \cite{Barnes2016} also calculated the  \ion{Na}{i} D lines' absorption depth and their result is 0.45\% (average of Night 1 and Night 3) which is also larger than in W2015 but smaller than ours.
The differences are probably due to the data reduction method, however, these differences are not significant considering the measurement uncertainties. When fitting the \ion{Na}{i} absorption, we use light curve models which include the limb-darkening within the bandwidths while the other two papers used a box light curve model.

As for the CLV effect, there is no obvious CLV feature in the light curve in W2015 (Fig.~3 of their paper), while the light curve in \cite{Barnes2016} (Fig.~6 of their paper) shows a similar CLV feature to that in our work. 
In general, the light curves in \cite{Barnes2016} and our work agree more closely than with W2015.

%                                                two column figure
%----------------------------------------------------------- S_vib

%_____________________________________________________________
\section{A Comparison of the CLV effect for HD 189733b and HD 209458b}
The actual CLV effect varies between different exoplanet systems. In order to illustrate and study the different manifestations of the CLV effect, we present the CLV of the HD 209458b system and compare it with the HD 189733b system. These two transiting exoplanet systems have bright host stars and have been extensively studied, thus they serve as typical samples for the study the CLV effect.

\begin{figure*}
   \centering
   \includegraphics[width=0.98\textwidth]{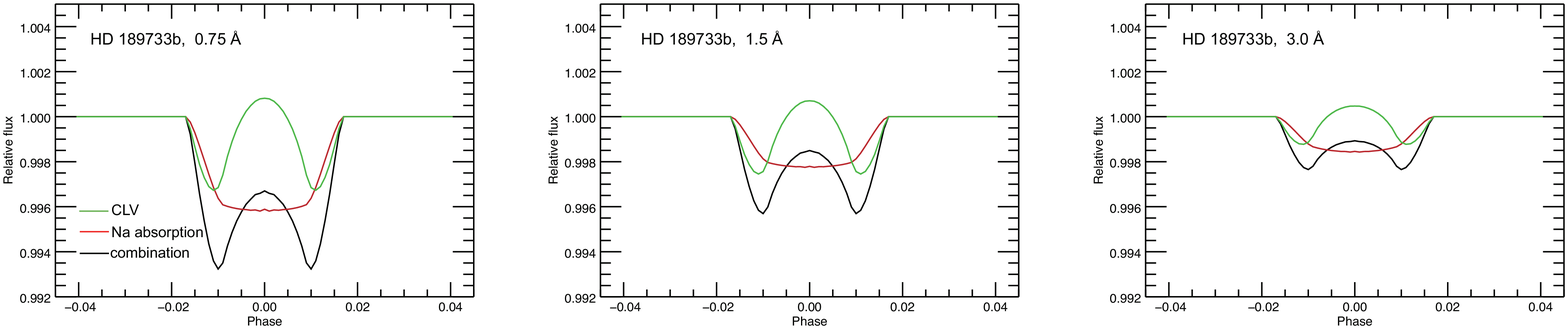}
   \includegraphics[width=0.98\textwidth]{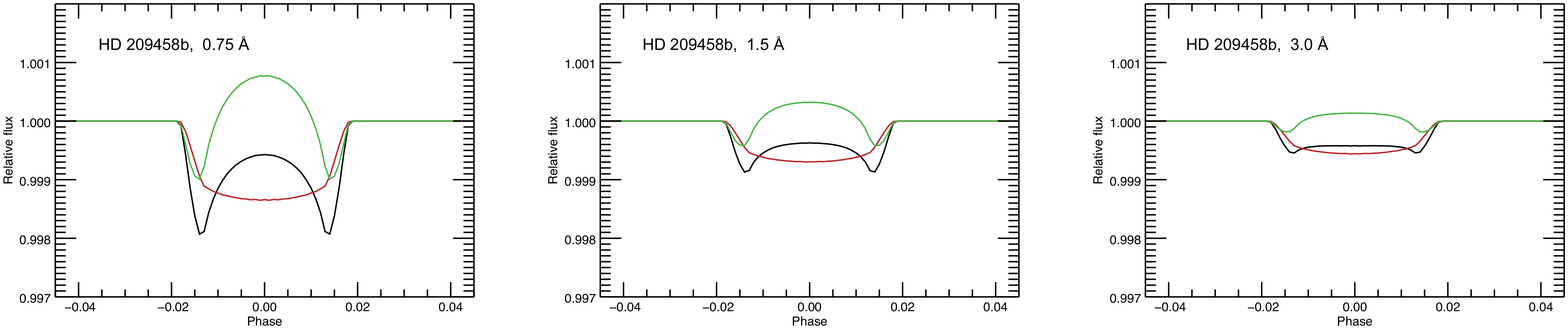}
\caption{Comparison of the CLV effect between HD 189733b and HD 209458b. Here the calculations are for the \ion{Na}{i} D lines (average of  $\mathrm{D_1}$ and  $\mathrm{D_2}$). The first row is for HD 189733b and the second row is for HD 209458b. The three columns are for the bandwidths of 0.75$\,\AA$, 1.5$\,\AA$ and 3.0$\,\AA$. The green lines are the modeled CLV light curves while the red lines are the \ion{Na}{i} absorption light curves, and the black lines are the combination of the two. Note the y axis scales are different for the two rows: the values for HD 189733b are significantly larger than for HD 209458b.
}
         \label{compare-HD209-HD189}
   \end{figure*}

We model the CLV effect of HD 209458b using the parameters shown in Table.\ref{parameter-HD209}. 
In order to compare the strengths of the CLV effect with the \ion{Na}{i} absorption, we adopt the measured planetary absorption values from \cite{Snellen2008} (S2008). The results are shown in Fig. \ref{compare-HD209-HD189} and Table. \ref{Table-Na-HD209}. 
Note that the calculation in S2008 does not correct for the CLV effect. However, since the overall CLV contribution is relatively small (the value is at the noise level of their observation data), adopting the S2008's value as the \ion{Na}{i} absorption is justified in our study.

%
%                                             Simple A&A Table
%-----------------------------------------------------------
\begin{table*}
\caption{Same as Table. \ref{Table-Na-HD189} but for HD 209458b. Here the measured \ion{Na}{i} absorption depths are from \cite{Snellen2008}. }             % title of Table
\label{Table-Na-HD209}      % is used to refer this table in the text
\centering                          % used for centering table
\begin{tabular}{c c c c c c }        % centered columns (4 columns)
\hline\hline                 % inserts double horizontal lines
Passband ($\,\AA$)	&	 Planetary absorption depth (\%)	 &	~ & CLV effect model  (\%) & ~ & CLV overall-transit /  \\     % table heading
	~ & (from S2008) & mid-transit & limb-transit & overall-transit & planetary absorption \\
	\hline                       % inserts single horizontal line
 		0.75	& 0.135	$\pm$ 0.017&	-0.077  & 0.099 & -0.007 & -5.3\%  \\
        1.5 	& 0.070	$\pm$ 0.011&	-0.032  & 0.042 & -0.004 & -5.7\%  \\
        3.0		& 0.056	$\pm$ 0.007&	-0.014  & 0.018 & -0.002 & -3.6\% \\
\hline                                   %inserts single line
\end{tabular}
\end{table*}
%-----------------------------------------------------------

%                                                two column figure
%----------------------------------------------------------- S_vib

\subsection{The differences}
In order to compare the two systems, we plot their CLV models together with the Na absorption at three different bandwidths in Fig. \ref{compare-HD209-HD189}. There are several sgnificant differences between the CLV effects of the two planetary systems.

\begin{figure}
   \centering
   \includegraphics[width=0.50\textwidth]{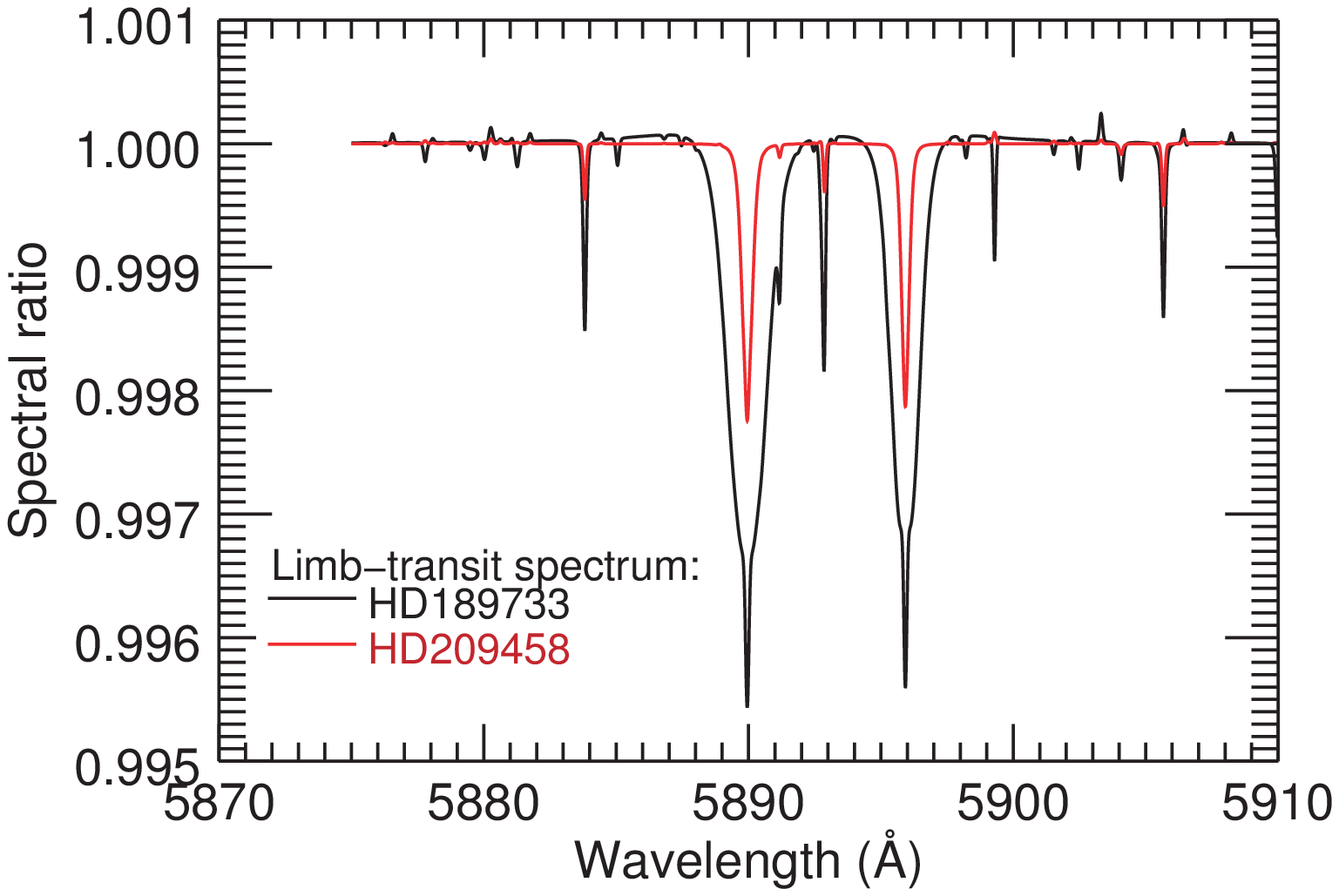}
\caption{The ratio of limb-transit spectrum to out-of-transit spectrum. The figure shows clearly that the CLV feature of HD 189733b is stronger and broader than the CLV feature of HD 209458b.
}
         \label{compare-limb-transit}
   \end{figure}

\begin{figure}
   \centering
   \includegraphics[width=0.50\textwidth]{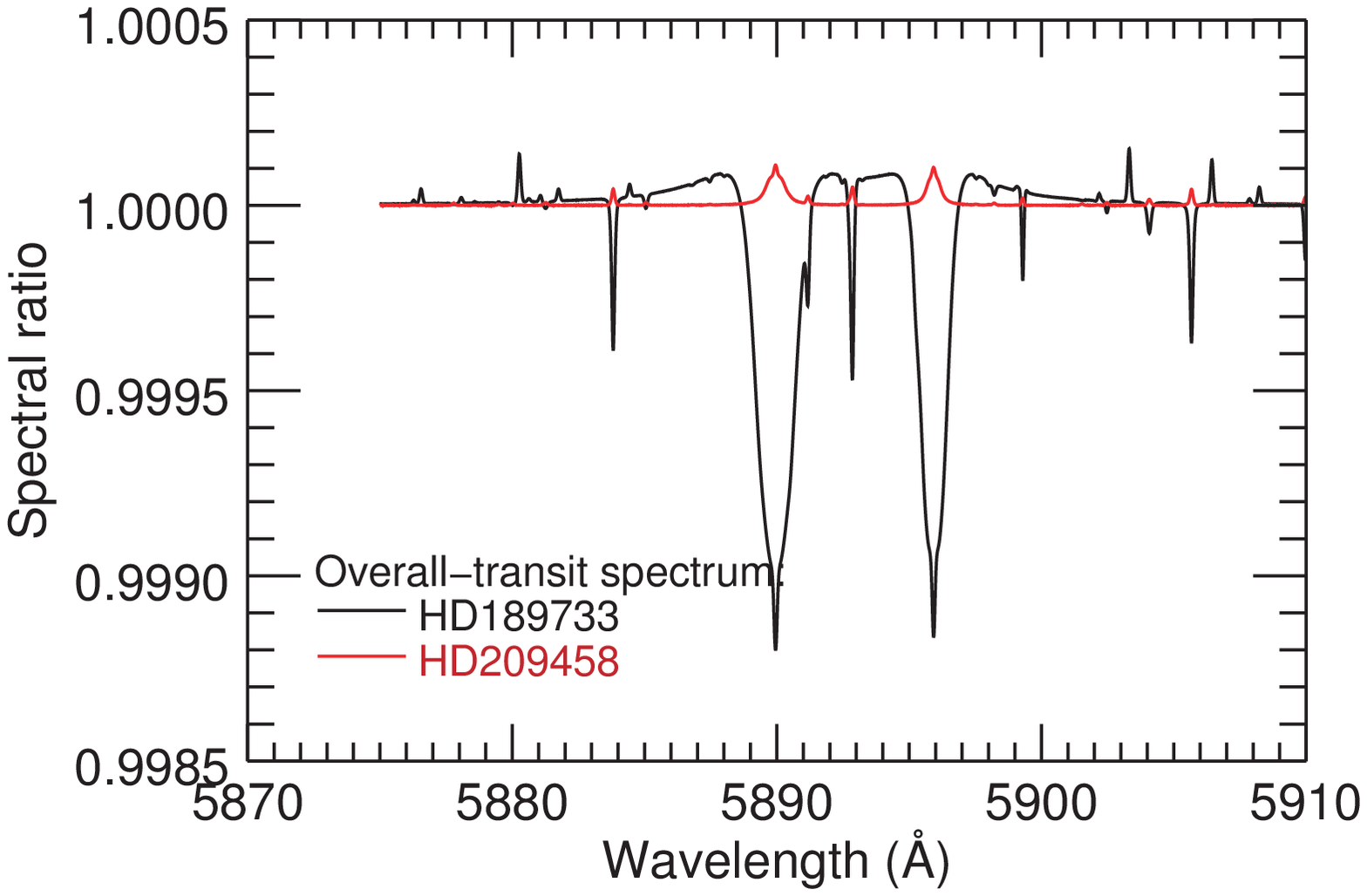}
\caption{The ratio of the overall-transit spectrum to the out-of-transit spectrum. Here the overall-transit spectrum is the combination of all the in-transit spectra.
This figure shows a significant difference between the overall-transit spectra of HD 189733b and HD 209458b and this difference is mainly attributed to their different transit parameters.
}
         \label{compare-overall-transit}
   \end{figure}

	\begin{enumerate}	%define item

	\item The first difference is in the strengths of the CLV effect. By comparing the CLV values in Table \ref{Table-Na-HD189} and \ref{Table-Na-HD209} and Fig.~\ref{compare-HD209-HD189}, we conclude that the CLV effect for HD 189733b is significantly larger than for HD 209458b. 
The difference is mainly attributed to the effective temperature of the host star (i.e. different spectral types). Generally a star with a lower effective temperature has a stronger CLV effect. The planet-to-star radius ratio also contributes to this difference, however, for these two exoplanet systems, this contribution is small considering that their planet-to-star radius ratios are similar (0.155 for HD 189733b and 0.124 for HD 209458b).

	\item The second difference is their behavior with different bandwidths. 
The CLV feature weakens with a broader bandwidth for both systems, however, the drop for HD 209458b is significantly larger than for HD 189733b. For example, the limb-transit value at 3.0 $\,\AA$ is 38\% of the value at 0.75 $\,\AA$ bandwidth for HD 189733b, while for HD 209458b the value is 18\%. To better demonstrate this difference, we plot the spectral ratio between limb-transit and out-of-transit for the two systems in Fig. \ref{compare-limb-transit}. From this figure, we can see that the CLV feature for HD 189733b extends over a larger wavelength range than for HD 209458b.

	\item The third difference is the overall-transit value (c.f. Table \ref{Table-Na-HD189} and \ref{Table-Na-HD209}). The overall-transit value is positive for HD 189733b (which means the line depth is deeper than out-of-transit), while the value is negative for HD 209458b (which means the line depth is shallower than out-of-transit). 
Their overall-transit spectra also show significant differences (Fig. \ref{compare-overall-transit}).
The reason is attributed to the transit impact parameter. 
A larger impact parameter means the planet obscures more limb positions on the stellar disk where the line depth is shallower than out-of-transit and thus the observed line depth is deeper, while a smaller impact parameter means the planet transits more centre positions where the line depth is deeper than out-of-transit and thus the observed line depth is shallower.
Since the impact parameter of HD 189733b (0.680) is larger than of HD 209458b (0.507), their overall-transit values are quite different.

   \end{enumerate}

%                                                one column figure
%----------------------------------------------------------- S_vib

%                                                one column figure
%----------------------------------------------------------- S_vib

%_____________________________________________________________
\section{General CLV effect during exoplanet transit}
As discussed in \cite{Yan2015}, there are a variety of parameters that affect the CLV feature, including stellar parameters and planet transit parameters. 
In this section, we study two important parameters: the stellar effective temperature and the transit impact parameter.

\subsection{Stellar effective temperature}
\begin{figure}
   \centering
   \includegraphics[width=0.50\textwidth]{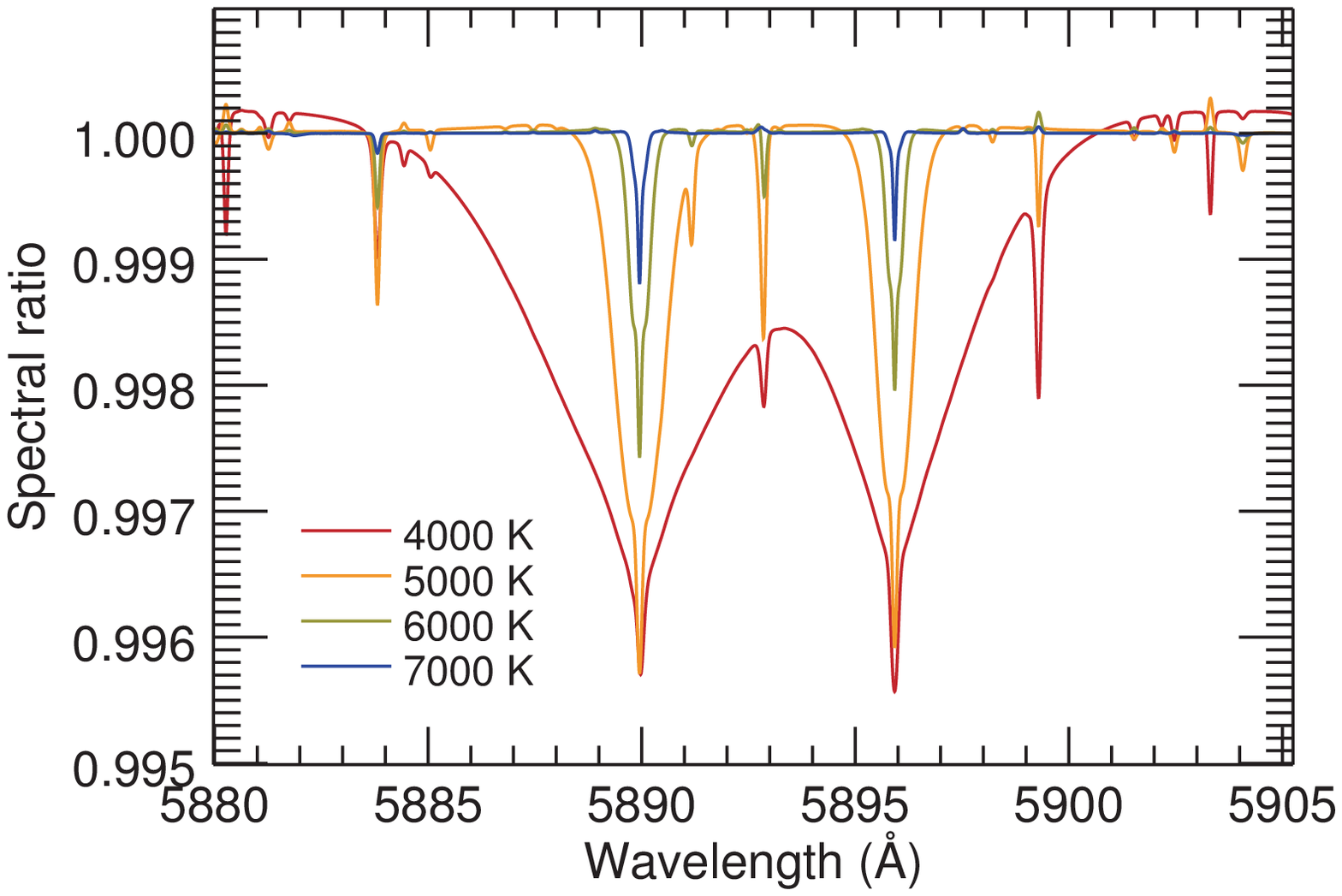}
\caption{The ratios of limb-transit spectra to out-of-transit spectra for stars with different $T_{\mathrm{eff}}$. These spectra are simulated using the planetary and stellar parameters described in Section 5.1.
Here, the limb-transit is defined as the orbital phase with the deepest line depth measured by using a 1.5 $\AA$ band centered at  the $\mathrm{D_2}$ line. This figure presents the change of the CLV effect with $T_{\mathrm{eff}}$.
}
         \label{Teff}
   \end{figure}

The intrinsic variations of stellar lines from the center to limb are determined by the stellar parameters. Among different stellar parameters, the effective temperature has the most significant impact on the CLV feature. 
The transit light curve of the \ion{Na}{i} D line varies significantly with $T_{\mathrm{eff}}$ (c.f. Fig.~5 and Table 2 in C2015). To demonstrate the importance of $T_{\mathrm{eff}}$, we construct the CLV effect models from 4000 K to 7000 K with a step of 1000 K.
We fix the following parameters in  the CLV model: stellar log($g$) (4.5), stellar metallicity ( [M/H]=0 ), impact parameter ($b=0$), planet to star radius ratio (0.1546 -- same as of HD 189733b). The modeled limb-transit spectra (ratios to the out-of-transit spectra) are plotted in Fig.~\ref{Teff}. 
This figure shows clearly that the CLV effect for the \ion{Na}{i} D lines  is stronger for stars with lower $T_{\mathrm{eff}}$.

%                                                one column figure
%----------------------------------------------------------- S_vib

The stellar effective temperature also affects the manifestation of the CLV effect with different bandwidths. 
Generally, the CLV value drops with a larger bandwidth, but there are differences between different spectral types. As can been seen in Fig.~\ref{Teff}, stars with low effective temperatures have relatively broad CLV features. Thus, for stars with low $T_{\mathrm{eff}}$, the CLV value drops less prominently with increasing bandwidth than for stars with high $T_{\mathrm{eff}}$.

\subsection{Transit impact parameter}
The planetary parameters are also important for the CLV effect, e.g. the planet radius and transit impact parameter. The effect of the planet-to-star radius ratio is straightforward, i.e. a larger planet radius results in a stronger CLV feature. Thus we focus here on the impact parameter ($b$).

\begin{figure}
   \centering
   \includegraphics[width=0.50\textwidth]{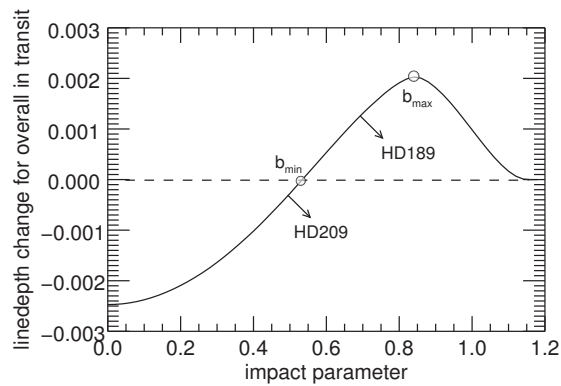}
\caption{The overall-transit values at different impact parameters (the `b-diagram'). The values are calculated using a 0.75 $\AA$ band centered at the \ion{Na}{i}  $\mathrm{D_2}$ line. The actual b values of HD 189733b and HD 209458b are also indicated. 
}
         \label{b-change}
   \end{figure}

\begin{figure}
   \centering
   \includegraphics[width=0.50\textwidth]{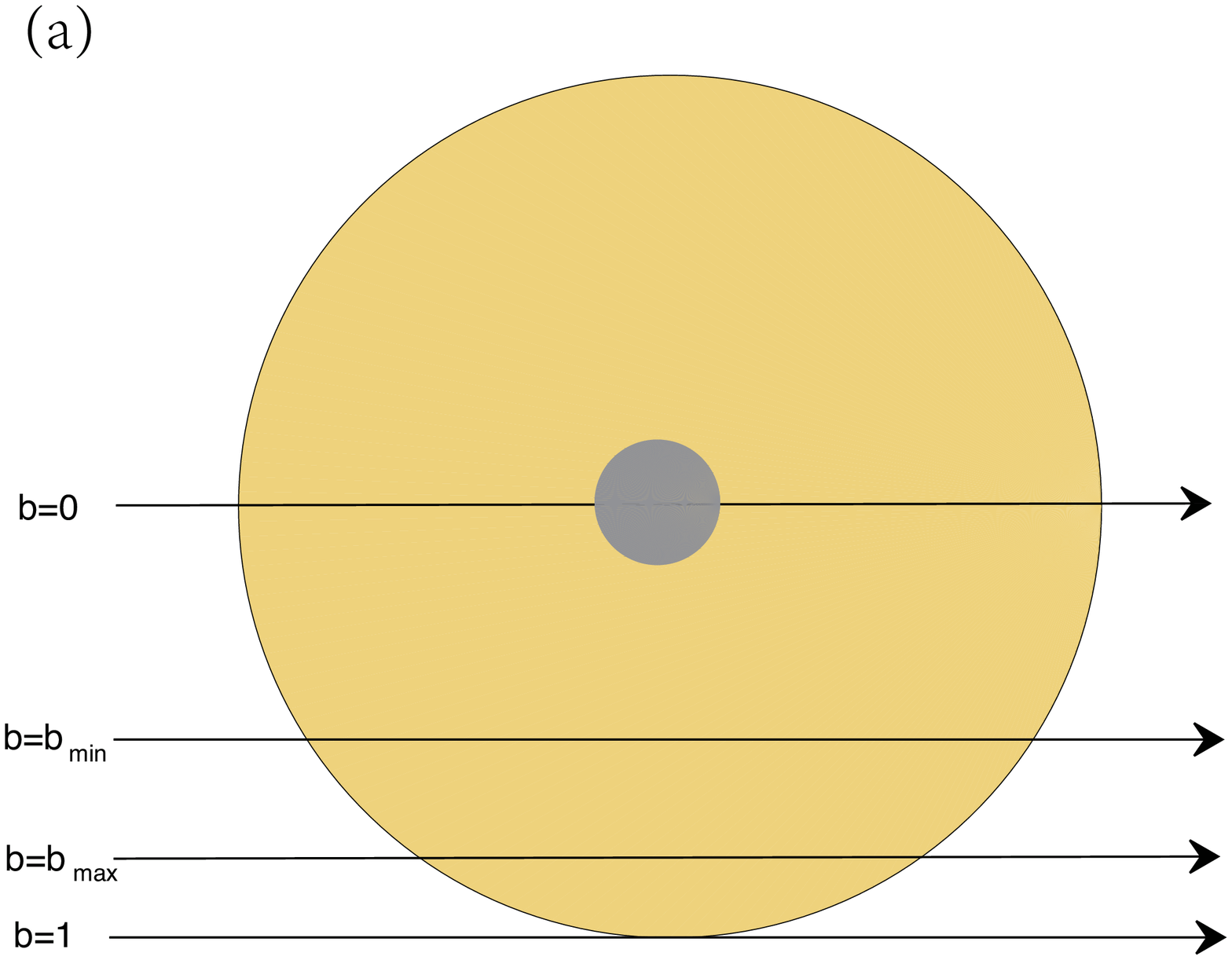}
   \includegraphics[width=0.50\textwidth]{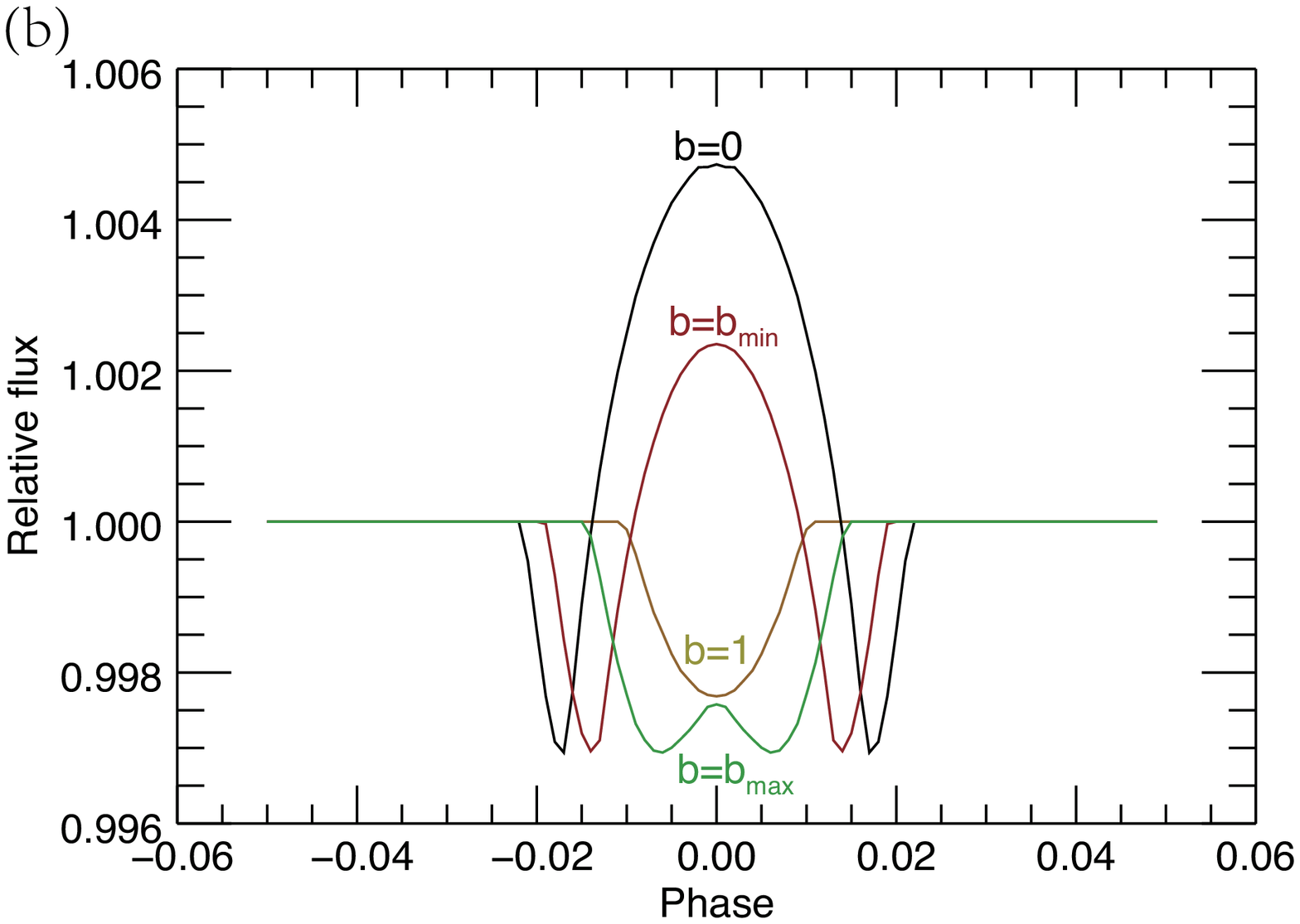}
\caption{\textit{(a)} The transit schematic with different impact parameters. \textit{(b)} The simulated CLV light curves of the \ion{Na}{i}  $\mathrm{D_2}$ line with a 0.75 $\AA$ band.
}
         \label{schematic-different-b}
   \end{figure}

We use the parameters of the HD 189733b system except the $b$ value to model the CLV feature. Figure. \ref{b-change} shows the change of the overall-transit value with different impact parameters. Here, the overall-transit value is the average line depth change for all the in transit spectra measured with a 0.75$\AA$ bandwidth centered at the \ion{Na}{i} $\mathrm{D_2}$ line. A positive value means the relative flux is smaller and the line depth is deeper compared to out-of-transit (i.e. similar to a planetary absorption), while a negative value means the relative flux is larger and the line depth is shallower (i.e. opposite to a planetary absorption). 
Figure. \ref{schematic-different-b} shows the transit schematic with different $b$ values and the corresponding CLV light curves.

Below we describe the CLV effect at typical $b$ values:
\\ \textbf{(1)}  When $b=0$ (edge-on transit), the overall-transit value is negative and its absolute value is the largest. This means the CLV light curve has the most prominent `bump' feature around mid-transit (the black line in Figure.~\ref{schematic-different-b}b).
\\ \textbf{(2)} When $b=0.54$ ($b_\mathrm{min}$), the overall-transit value is zero, which means the average CLV effect during transit does not change the line depth. 
\\ \textbf{(3)} The largest overall-transit value appears at  $b=0.84$ ($b_\mathrm{max}$). This positive largest value means the CLV effect makes the line depth much deeper than out-of-transit. The CLV light curve at  $b_\mathrm{max}$ (the green line in Figure. \ref{schematic-different-b}b) is very similar to the planetary absorption light curve. Thus, when the impact parameter of an exoplanet transit is around $b_\mathrm{max}$, the CLV effect could mimic the signal of the atmospheric absorption.
\\ \textbf{(4)} When  $b=1$, the CLV effect is relatively weak as  the transit is grazing and only a small part of the stellar disk is obscured.

We also calculated the $b_\mathrm{max}$ and $b_\mathrm{min}$ values for other bandwidths (1.5 $\AA$ and 3.0 $\AA$) and the $\mathrm{D_1}$ line. The results deviate only slightly (about 0.02) from the values of $\mathrm{D_2}$ 0.75 $\AA$. So this b-diagram represents the general behaviour of the CLV effect at the \ion{Na}{i} D doublets.

For stars with different $T_{\mathrm{eff}}$, we model their b-diagrams using the same parameters described in Section 5.1. The calculated b-diagrams have the same shape as shown in Fig.~\ref{b-change}, only that the actual line depth values are more significant with a decreasing $T_{\mathrm{eff}}$.  The $b_\mathrm{max}$ and $b_\mathrm{min}$ values for 4000 $\sim$ 6000 K are all similar to those of HD 189733b (the deviations are smaller than 0.02). For a star with $T_{\mathrm{eff}}$=7000 K, the  $b_\mathrm{max}$ and $b_\mathrm{min}$ are 0.88 and 0.64 (for $\mathrm{D_2}$ 0.75 $\AA$), which are different from the values of HD 189733b.

We  also model the b-diagrams with different planet to star radius ratios ($R_\mathrm{p}/R_\mathrm{s}$). We change the $R_\mathrm{p}/R_\mathrm{s}$ from 0.01 to 0.2 and keep all the other parameters the same as the HD 189733b system. The  $b_\mathrm{min}$ value changes by less than 0.02 and the $b_\mathrm{max}$ value decreases from 0.90 to 0.82 with the increasing $R_\mathrm{p}/R_\mathrm{s}$. 

With this b-diagram, the overall-transit difference between HD 189733b and HD 209458b can be easily understood. The actual impact parameter of HD 189733b is 0.680 which is larger than $b_\mathrm{min}$ and thus the CLV effect makes the overall-transit line depth deeper. The impact parameter of HD 209458b is 0.507 which is smaller than but close to $b_\mathrm{min}$, so the CLV effect results in a shallower line depth and the effect is relatively weak.

%                                                one column figure
%----------------------------------------------------------- S_vib

%                                                one column figure
%----------------------------------------------------------- S_vib

\subsection{Stellar atmosphere model}
The modeled CLV feature depends on the stellar atmosphere model. Normally a one-dimensional LTE stellar model can produce a reasonably realastic CLV feature, however, non-LTE and three-dimensional (3D) models can produce more precise results.
For the Sun, \cite{Allende2004} used the CLV of solar lines to test the non-LTE calculations and \cite{Koesterke2008} studied the solar CLV with a 3D model. Recently, \cite{Dravins2016} used the transit spectrum of HD 209458b to study the CLV of \ion{Fe}{i} lines with a 3D model.

\begin{figure}
   \centering
   \includegraphics[width=0.50\textwidth]{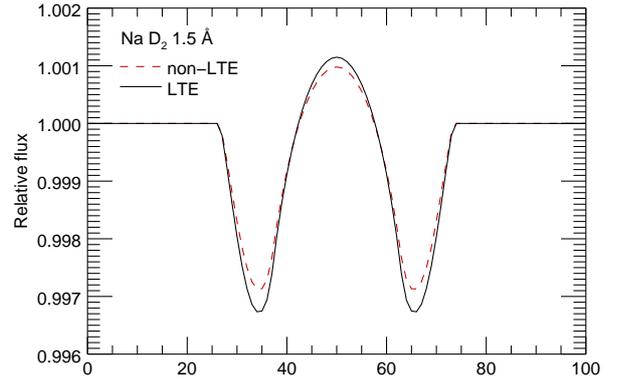}
\caption{The CLV light curves of $\mathrm{D_2}$ line from LTE and non-LTE calculations. Here the simulations use the parameters of the HD 189733b system.
}
         \label{NLTE}
   \end{figure}

In our work, we employ a one-dimensional non-LTE model using the SME tool. Compared to the LTE calculation, the non-LTE calculation results in a less prominent CLV effect for the \ion{Na}{i} D lines. Fig.~\ref{NLTE} shows the comparison between the LTE and non-LTE cases. The CLV light curves in the figure are calculated for the $\mathrm{D_2}$ line (1.5 $\AA$ band) using the parameters of the HD 189733b system. The same Na abundance value is used in the LTE and non-LTE calculations.

%                                                one column figure
%----------------------------------------------------------- S_vib

\subsection{CLV effect for low-mass stars}
\begin{figure}
   \centering
   \includegraphics[width=0.50\textwidth]{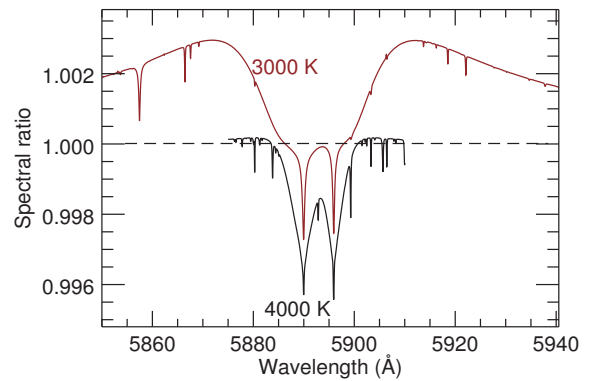}
\caption{The spectral ratios of limb-transit to out-of-transit for a low-mass star with  $T_{\mathrm{eff}}$=3000 K. The spectral ratio for a star with $T_{\mathrm{eff}}$=4000 K is also plotted for comparison. One can see a different behavior of the line wing for 3000 K. The wavelength range here is larger than in Fig.~\ref{Teff} in order to give a clearer view of the broad line profile.
}
         \label{Teff-3000K}
   \end{figure}

In Section 5.1, we modeled the CLV effects for stars with $T_{\mathrm{eff}}$ ranging from 4000 K to 7000 K. However, low-mass stars with lower $T_{\mathrm{eff}}$  are very interesting targets for exoplanet atmosphere studies given their low masses and small radius which make the planet detection and atmospheric characterisation easier. For instance, the recently discovered terrestrial exoplanets around TRAPPIST-1 \citep{Gillon2016} and Proxima Centauri \citep{Anglada2016} are both orbiting M-type stars. The CLV effect for low-mass stars is different compared to other spectral types. We synthesize the stellar spectrum for low-mass stars using the SME tool with the MARCS2012 stellar model (which has a  $T_{\mathrm{eff}}$ grid from 2500 K to 8000 K) and the VALD line list (which is retrieved with a stellar $T_{\mathrm{eff}}$ of 3500 K). The stellar model and line list are valid for low $T_{\mathrm{eff}}$.
In Fig.~\ref{Teff-3000K}, we present the result for a star with $T_{\mathrm{eff}}$=3000 K and log($g$)=5.2 (all other parameters are the same as in Section 5.1). 
The limb-transit spectrum for a low-mass star has a different behaviour between line core and line wing. As shown in the figure, at the line wing with wavelength smaller than 5886 $\AA$ or larger than 5898 $\AA$, the line depth for the limb-transit spectrum is shallower than for out-of-transit. While in the line core regions, the CLV feature is similar to other spectral types.
The CLV effect of low-mass stars is particularly important because the \ion{Na}{i} D lines are very broad and deep and have a relatively complicated wavelength-dependent CLV feature.

%                                                one column figure
%----------------------------------------------------------- S_vib

%_____________________________________________________________
\section{Conclusions}
We have used the HARPS archive data of HD 189733b transits to show the CLV effect for the \ion{Na}{i} D lines. The \ion{Na}{i} transmission light curves are the combination of the CLV effect and the actual planetary absorption. After the correction of the CLV effect, the transmission light curve fits the planetary absorption better and the measurement of \ion{Na}{i} absorption becomes more precise.
We model the CLV effect and the Na absorption of HD 209458b and HD 189733b and compare the CLV effects of these two benchmark exoplanet systems. Due to the different effective temperatures of their host stars, the CLV effect for HD 189733b is much stronger than for HD 209458b, and the CLV feature of HD 189733b extends to a lager wavelength band than that of HD 209458b. The overall-transit contribution (the average of all the in-transit spectra) of the CLV effect behaves oppositely for the two planetary systems, i.e. the overall-transit spectrum compared to the out-of-transit spectrum is deeper for HD 189733b while shallower for HD 209458b. This is attributed to their different transit impact parameters.

We further model the CLV effect for planets orbiting stars with various effective temperatures. In general, the \ion{Na}{i} D line's CLV effect is stronger and has a broader wavelength range for stars with lower $T_{\mathrm{eff}}$.
For low-mass stars with very low $T_{\mathrm{eff}}$, the CLV effect are very significant and thus important for transmission spectroscopy. However, spectral modeling for very low $T_{\mathrm{eff}}$ stars is relatively complicated and further work is needed.

The transit impact parameter ($b$) is important for the CLV effect. The $b$ value determines the trajectory of the planet transit and so the actual CLV effect varies with $b$. 
We introduce a b-diagram which shows the change of the overall-transit value with  $b$ (c.f. Fig.~\ref{b-change}). When $b=0$, the overall-transit value for the \ion{Na}{i} D line is negative, which means the stellar line in the overall-transit spectrum is shallower than in the out-of-transit spectrum. 
At a particular $b$ value ($b_\mathrm{min}$) where the overall-transit is zero, the net CLV effect has no influence on the stellar line depth.
When $b$ = $b_\mathrm{max}$ (where the overall-transit value is the largest), the planet transits mostly the limb part of the stellar disk and the line depth is much deeper than in out-of-transit. For an impact parameter close to $b_\mathrm{max}$,  the CLV effect has a similar light curve to an exoplanet atmosphere absorption. 
We calculate the b-diagrams of  \ion{Na}{i} D lines for stars with $T_{\mathrm{eff}}$  ranging from 4000 K to 6000 K. According to our simulations, the $b_\mathrm{max}$ is generally around 0.84 and $b_\mathrm{min}$ is around 0.54, with deviations smaller than 0.02.

With the improvement of instrumentation (e.g. future instruments like ESPRESSO on VLT, HIRES on E-ELT), transit spectroscopy observations will be applied to many exoplanets and the transmission spectra will have a high signal-to-noise ratio. Thus, the CLV effect will become crucial for species detection as well as for the determination of physical conditions using transit spectroscopy.

Future work to improve the CLV model is necessary as the transmission observation is becoming more precise. These investigations will benefit from the employment of 3D stellar models, a proper treatment of the non-LTE effects and an improved modeling for stars with very low  $T_{\mathrm{eff}}$.

\begin{acknowledgements}
We would like to thank Yeisson Osorio, Hans-Günter Ludwig and Karin Lind for helpful discussions of stellar spectral models. We thank the referee for helpful comments. This research is based on data obtained from the ESO Science Archive Facility under request number 217581 and 219943. The observation is made with the HARPS instrument on the ESO 3.6-m telescope at the La Silla Observatory under the programme ID 072.C-0488(F) and 079.C-0127(A). This work has made use of the VALD database, operated at Uppsala University, the Institute of Astronomy RAS in Moscow, and the University of Vienna.
\end{acknowledgements}

% for the bibliography, at the end
\bibliographystyle{aa} % style aa.bst

\bibliography{CLVpaper-2016}

\end{document}